\documentclass{article}
\usepackage{arxiv}

\usepackage[utf8]{inputenc} 
\usepackage[T1]{fontenc}    
\usepackage{hyperref}       
\usepackage{url}            
\usepackage{booktabs}       
\usepackage{amsfonts}       
\usepackage{nicefrac}       
\usepackage{microtype}      

\usepackage{amsmath,amsfonts,amssymb,graphicx}

\title{pylustrator: code generation for reproducible figures for publication}

\author{
  Richard C.~Gerum \\
  Department of Physics\\
  Friedrich Alexander Universität Erlangen-Nürnberg\\
  Erlangen, Germany \\
  \texttt{richard.gerum@fau.de}
}

\begin{document}

\maketitle

\begin{abstract}
One major challenge in science is to make all results potentially reproducible. Thus, along with the raw data, every step from basic processing of the data, evaluation, to the generation of the figures, has to be documented as clearly as possible. While there are many programming libraries that cover the basic processing and plotting steps (e.g.\ Matplotlib in Python), no library yet addresses the reproducible composing of single plots into meaningful figures for publication. Thus, up to now it is still state-of-the-art to generate publishable figures using image-processing or vector-drawing software leading to unwanted alterations of the presented data in the worst case and to figure quality reduction in the best case.
\textit{Pylustrator} a open source library based on the Matplotlib aims to fill this gap and provides a tool to easily generate the code necessary to compose publication figures from single plots. It provides a graphical user interface where the user can interactively compose the figures. All changes are tracked and converted to code that is automatically integrated into the calling script file. Thus, this software provides the missing link from raw data to the complete plot published in scientific journals and thus contributes to the transparency of the complete evaluation procedure.

\end{abstract}

\section{Introduction}

In recent years, more and more researchers have called attention to a growing "reproducibility crisis" \cite{CRL16846}. An important factor that contributes to problems in reproducing results from published studies is the unavailability of the raw data from the original experiment as well as the unavailability of the methods or the code used for the evaluation the raw data \cite{Baker2016}. 
One major step to overcome these shortcomings is the publication of all raw data as well as a documented version of the code used for evaluation \cite{baker2016scientists}. The ideal case would be that anyone interested can download the raw data and exactly reproduce the figures of the publication.

To address the issue of data availability, researchers are encouraged to provide their data in online repositories, e.g. dryad \cite{dryad}. However, this data is useless unless the complete evaluation procedure in the terms of all evaluation and visualisation steps
can be comprehended by other scientists. The best way to to so is to provide a complete well documented evaluation code, 
including all important steps from basic artifact corrections up to the final plot to be published. Open Source scripting languages like Python \cite{Rossum1995} or R \cite{R} are ideal for such code as open source languages are accessible for everyone. Furthermore, interpreted languages do not need to be compiled, therefore have less obstacles for the user to run the code. The last part of the evaluation of the data is the visualisation, which is crucial to communicate the results \cite{Tufte1893}. 
This paper deals with the visualization step which consists of two parts: the generation of simple plots from data and composing meaningful figures from these plots.

The first part of generating the building blocks of figures, the plots, is already covered in various toolkits, e.g. Matplotlib \cite{Hunter2007}, Bokeh \cite{Bokeh} or Seaborn \cite{seaborn}. But to generate reproducible figures from simple plots, no convenient toolkit is yet available. Matplotlib already offers figures composed of multiple subplots, but to generate a complete figure ready for publication a lot of code is needed to add all formatting, annotations and styling commands. Therefore, this approach is often not followed as it is impractical for real applications. Users often prefer graphical tools such as image manipulation software, e.g. GIMP \cite{GIMP} or Inkscape \cite{Inkscape}. These offer great flexibility, but cannot provide a reproducible way of generating figures and bear the danger of accidentally changing data points. Also important to note is that by using an image manipulation software, any small change to the evaluation requires to re-edit the figure in the image manipulation software. A process that slows down the creation of figures and is prone to errors.

\section{Algorithm and Exemplary Results}

\begin{figure}[htbp]
\centering
\includegraphics[width=0.8\textwidth]{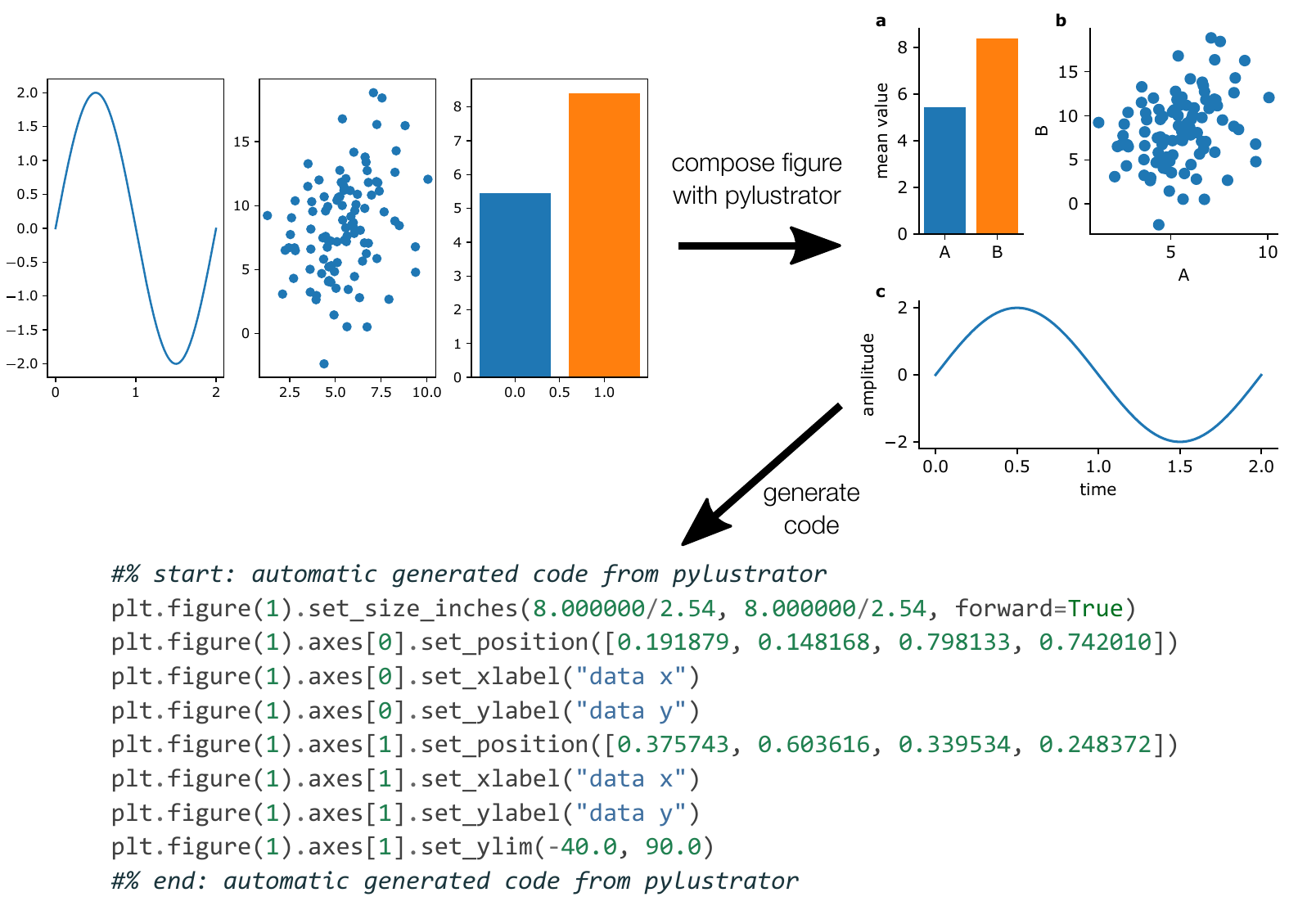}
\caption{Example how code for composing a figure can be generated with pylustrator.}
\label{fig:Visualisation}
\end{figure}

\textit{Pylustrator} was developed to address this issue. A tool to fill the gap from single plots to complete figures, by a code generation algorithm, that turns user input into python code for reproducible figure assembly (Fig.~\ref{fig:Visualisation}).  Small changes to the evaluation or new data only require to run the code again to update the figure.

\begin{figure}[htbp]
\centering
\includegraphics[width=\textwidth]{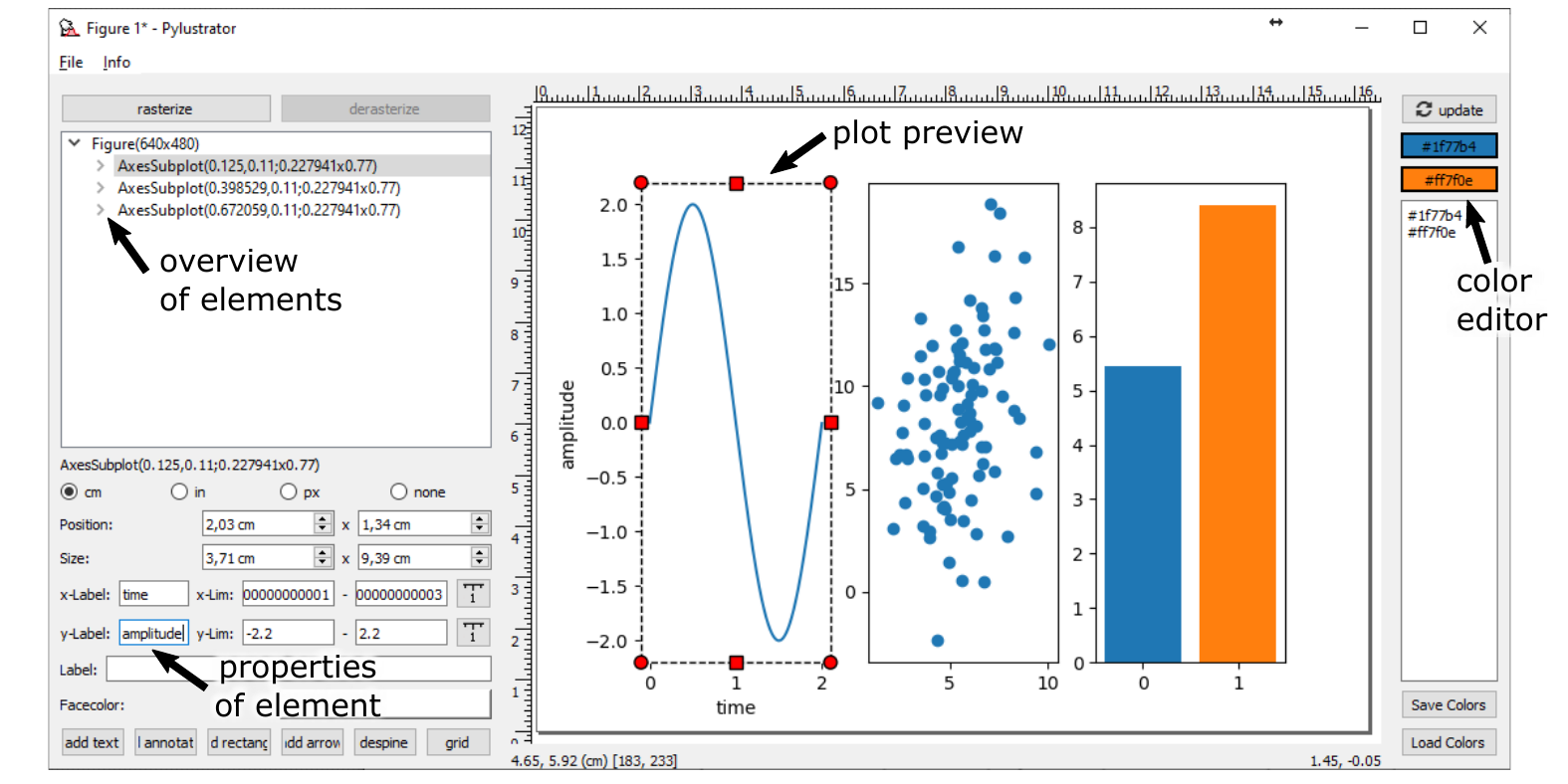}
\caption{The interface of \textit{pylustrator}. The user can view the elements of the plot, edit their properties, edit them in the plot preview and experiment with different color schemes.}
\label{fig:Interface}
\end{figure}

Using ``pylustrator`` in any given Python file that uses Matplotlib do plot data, simply requires the addition of only two lines of code:

\begin{verbatim}
    import pylustrator
    pylustrator.start()
\end{verbatim}

The Matplotlib figure is then displayed in an interactive window (Fig.~\ref{fig:Interface}) when the \texttt{plt.show()} command is called. In this interactive window, \textit{pylustrator} enables the user to:

\begin{itemize}
\item resize and position plots by mouse-dragging 
\item adjust the position of plots legends
\item align elements easily by automatic "snapping"
\item resize the complete figure in cm/inches
\item add text and annotations, and change their style and color
\item adjust plot ticks and tick labels
\end{itemize}

\textit{Pylustrator} tracks all these changes to translate them into python code. Every change is split in four parts: the command object, the command text, the target object and the target command. The command object is the object instance (e.g. the Axes object) that has a method to call for this change and the command text is the methods name together with the parameters (e.g. ".annotate('New Annotation')"). The target object is the object that is affected by the command. In most cases this is the same as the command object, but in some cases when new child objects are created the target object is the child object. The target command is the methods name without the parameters.
 
 Command objects are "serialized" by iteratively going up the parent-child tree from e.g. a text to the axis to the figure and generating a python command from this dependency (e.g. `plt.figure(1).axes[0].texts[0]`, the first text of the first axes of figure 1). When saving, \textit{pylustrator} introspects its current stack to find the line of code from where it was called and inserts the automatically generated code directly before the command calling \textit{pylustrator}.
 
 When loading a file with automatically generated code, \textit{pylustrator} splits all the automatically generated lines into reference objects and reference commands. New changes where both the reference object and the reference command match a previous change, the previous change is overwritten. This ensures that previously generated code can be loaded appropriately and saving the same figure multiple times generates the code only once.
  
It is important to note that the automatically generated code only relies on Matplotlib and does not need the \textit{pylustrator} package anymore. Thus, the \textit{pylustrator} import can later be removed to allow to share the code without an additional introduced dependency. 

The documentation of \textit{pylustrator} can be found on \url{https://pylustrator.readthedocs.org}.

\section{Conclusion}
This study introduces a novel method to create publishable figures from single plots based on an open source Python library called \textit{pylustrator}. The figures can be arranged by drag and drop and the \textit{pylustrator} library produces the according code. Thus, this library provides a valuable contribution to tackle the reproducibility crisis.

\section{Acknowledgements} 

We acknowledge testing, support and feedback from Christoph Mark, Sebastian Richter, and Achim Schilling and Ronny Reimann for the design of the Pylustrator Logo.

\bibliographystyle{unsrt}
\bibliography{paper}

\begin{thebibliography}{10}

\bibitem{CRL16846}
Franklin Sayre and Amy Riegelman.
\newblock The reproducibility crisis and academic libraries.
\newblock {\em College \& Research Libraries}, 79(1):2, 2018.

\bibitem{Baker2016}
Monya Baker and Dan Penny.
\newblock {Is there a reproducibility crisis?}, may 2016.

\bibitem{baker2016scientists}
Monya Baker.
\newblock Why scientists must share their research code.
\newblock {\em Nature News}, 2016.

\bibitem{dryad}
Dryad.
\newblock \url{https://datadryad.org}.
\newblock Accessed: 2019-09-19.

\bibitem{Rossum1995}
Guido {Van Rossum} and Fred~L {Drake Jr}.
\newblock {\em {Python tutorial}}.
\newblock Centrum voor Wiskunde en Informatica Amsterdam, The Netherlands,
  1995.

\bibitem{R}
{R Core Team}.
\newblock {\em R: A Language and Environment for Statistical Computing}.
\newblock R Foundation for Statistical Computing, Vienna, Austria, 2019.

\bibitem{Tufte1893}
Edward Tufte.
\newblock {\em {The visual display of quantitative information}}.
\newblock Graphics Press, Cheshire, Connecticut, 1893.

\bibitem{Hunter2007}
J~D Hunter.
\newblock {Matplotlib: A 2D graphics environment}.
\newblock {\em Comput. Sci. Eng.}, 9(3):90--95, 2007.

\bibitem{Bokeh}
{Bokeh Development Team}.
\newblock {\em Bokeh: Python library for interactive visualization}, 2019.

\bibitem{seaborn}
Michael Waskom, Olga Botvinnik, Drew O'Kane, Paul Hobson, Saulius Lukauskas,
  David~C Gemperline, Tom Augspurger, Yaroslav Halchenko, John~B. Cole, Jordi
  Warmenhoven, Julian de~Ruiter, Cameron Pye, Stephan Hoyer, Jake Vanderplas,
  Santi Villalba, Gero Kunter, Eric Quintero, Pete Bachant, Marcel Martin, Kyle
  Meyer, Alistair Miles, Yoav Ram, Tal Yarkoni, Mike~Lee Williams, Constantine
  Evans, Clark Fitzgerald, Brian, Chris Fonnesbeck, Antony Lee, and Adel
  Qalieh.
\newblock mwaskom/seaborn: v0.8.1 (september 2017), September 2017.

\bibitem{GIMP}
{GIMP Development Team}.
\newblock {\em GIMP:GNU Image Manipulation Program}, 2019.

\bibitem{Inkscape}
The~Inkscape Team.
\newblock {\em Inkscape}, 2019.

\end{thebibliography}

\end{document}